\documentclass[aps,prc,preprint,amsmath,amssymb,showpacs,preprintnumbers,superscriptaddress]{revtex4-1}
\usepackage{CJK}
\usepackage{graphicx}
\usepackage{dcolumn}
\usepackage{bm}
\usepackage{slashed}
\usepackage{ulem}
\usepackage{epstopdf}
\usepackage{bbold}
\usepackage{mathrsfs}   
\usepackage{color,xcolor}  
\usepackage{multirow}  
\usepackage{booktabs} 

\usepackage{hyperref}

\newcommand{\bbm}{\begin{bmatrix}}
\newcommand{\ebm}{\end{bmatrix}}
\newcommand{\bBm}{\begin{Bmatrix}}
\newcommand{\eBm}{\end{Bmatrix}}
\newcommand{\bpm}{\begin{pmatrix}}
\newcommand{\epm}{\end{pmatrix}}

\begin{document}


\title{Hadron-quark phase transition in neutron star by combining the relativistic Brueckner-Hartree-Fock theory and Dyson-Schwinger equation approach}

\author{Pianpian Qin}
\affiliation{Department of Physics and Chongqing Key Laboratory for Strongly Coupled Physics, Chongqing University, Chongqing 401331, China}

\author{Zhan Bai}
\affiliation{Institute of Theoretical Physics, Chinese Academy of Sciences, Beijing 100190, China}

\author{Sibo Wang}
\affiliation{Department of Physics and Chongqing Key Laboratory for Strongly Coupled Physics, Chongqing University, Chongqing 401331, China}

\author{Chencan Wang}
\affiliation{School of Physics and Astronomy, Sun Yat-Sen University, Zhuhai 519082, China}

\author{Si-xue Qin}
\email{sqin@cqu.edu.cn}
\affiliation{Department of Physics and Chongqing Key Laboratory for Strongly Coupled Physics, Chongqing University, Chongqing 401331, China}

\date{\today}

\begin{abstract}
Starting from the relativistic Brueckner-Hartree-Fock theory for nuclear matter and the Dyson-Schwinger equation approach for quark matter, the possible hadron-quark phase transition in the interior of a neutron star is explored.
The first-order phase transition and crossover are studied by performing the Maxwell construction and three-window construction respectively.
The mass-radius relation and the tidal deformability of the hybrid star are calculated and compared to the 
joint mass-radius observation of a neutron star and the constraints from gravitational wave detection.
For the Maxwell construction, no stable quark core is found in the interior of a neutron star.
For the three-window construction, the parameters of the smooth interpolation function are chosen in such a way to keep the thermodynamic stability and lead to a moderate crossover density region.
To support a two-solar-mass neutron star under the three-window construction, the effective width of medium screening effects in quark matter should be around $0.35$~GeV.

\end{abstract}

\maketitle


\section{Introduction}\label{SecI}

At sufficiently high temperature or density,
the quarks are likely to ``escape'' from nucleons and become basic degrees of freedom.
Such transition is known as the hadron-quark phase transition.
Unlike the transition at high temperature, which is reachable in heavy ion collisions,
the transition at large density is far beyond the scope of terrestrial experiments.
The corresponding density can only be found in astronomical compact objects, such as neutron stars.
The neutron star is the remnant of a supernova explosion,
and the density inside a neutron star could reach about $5\sim 10\rho_{\text{sat}}$,
where $\rho_{\text{sat}}=0.16~\text{fm}^{-3}$ is the nuclear saturation density~\cite{2004-Lattimer-science.1090720,2021-Lattimer-ARNPS}.
At such a density, it is very likely that the hadron-quark phase transition will take place and deconfined quark matter appears~\cite{2019-Annala-NaturePhysics,2019-Bauswein-PhysRevLett.122.061102}.
Therefore, it is possible to use an astronomical observation of neutron stars to constrain the theories about the dense matter and phase transition.


Currently, it is difficult to unitedly describe the hadron matter, quark matter and hadron-quark phase transition within a single theoretical framework.
To study the hadron-quark phase transition inside the hybrid star, 
it is common to describe the hadron and quark matter separately with corresponding approaches,
and then use different construction schemes to combine them and get a complete equation of state (EOS).

For the hadron matter, there have already been many nuclear many-body methods,
among which the nonrelativistic~\cite{2003-Stone-PhysRevC.68.034324,2013-Fattoyev-PhysRevC.87.015806} and relativistic~\cite{2018-Fattoyev-Phys.Rev.Lett.120.172702,2020-LiJJ-PLB,2022-FuHanRui-Phys.lett.B834.137470, 2008-Sun-PhysRevC.78.065805,Tong_2020-PhysRevC.101.035802}
density functional theories (DFTs) are very important.
They are based on effective nucleon-nucleon $(NN)$ interactions, 
where the parameters are determined by fitting the ground state properties of finite nuclei and infinite nuclear matter at saturation.
The predictions of the neutron star properties with different DFTs are rather divergent due to the loose constraints of the effective $NN$ interactions at higher densities.
In comparison, one can start from realistic $NN$ interactions where the parameters are constrained with the $NN$ scattering data in free space, 
and utilize the many-body methods to deal with the realistic $NN$ interactions, 
such as the relativistic Brueckner-Hartree-Fock (RBHF) theory~\cite{2019-Shen-PPNP} as well as its nonrelativistic counterpart BHF theory~\cite{2007-Baldo-JPG}.
In particular, the RBHF theory is rooted in the relativistic framework, which contains significant three-body force effects self-consistently~\cite{2012-Sammarruca-PhysRevC.86.054317} and naturally avoids the problem of superluminance in nonrelativistic methods~\cite{1992-GQLi-Phys.Rev.C}.
The RBHF theory has been successfully applied to study the neutron star with pure hadron matter~\cite{Krastev2006,CCWang_2020_ApJ,Tong2022ApJ}.

In the description of quark matter, phenomenological models are widely used
(see, e.g., Refs.~\cite{1994-Hatsuda-pr,1999-Buballa-plb,2004-Fukushima-plb,1995-Benvenuto-PhysRevD.51.1989,2013-Torres-epl,1974-Chodos-PhysRevD.9.3471,2015-Zacchi-PhysRevD.92.045022}). 
In spite of these successful models, a study that is directly based on quantum chromodynamics (QCD) is still needed.
The most important first-principle method, the lattice QCD~\cite{1997-Montvay,2019-Ding-PhysRevLett.123.062002},
is powerful at zero chemical potential but encounters the notorious “sign problem” at large density. 
The perturbative QCD~\cite{Kurkela:2009gj,Kurkela:2016was} also loses its power in the phase transition region.
Therefore, a nonperturbative, continuum approach is required to study the phase transition of cold-dense matter.
In particular, the Dyson-Schwinger equation (DSE) approach~\cite{Roberts2000PPNP,Fischer2019PPNP} is a typical functional method based on QCD. 
It can deal with the confinement and dynamical chiral symmetry breaking simultaneously, 
and can be naturally used at finite temperature and chemical potential. 
In recent years, there have already been studies of neutron stars by using the DSE method
~\cite{2011-ChenH-PhysRevD.84.105023,2012-ChenH-PhysRevD.86.045006,2015-ChenH-PhysRevD.91.105002,2016-Yasutake-JPCS,2018-Bai-PhysRevD.97.023018,2021-Bai-EPJC}.
However, the relevant study is still in its early stages, and more efforts must be paid to make this method mature.

As for the construction of phase transition,
the most widely used method is the Maxwell construction~\cite{Endo:2006PTP,Hempel:2009PRD,Yasutake:2009PhysRevD.80.123009}.
It assumes that the phase transition is of first order~\cite{1992-Glendenning-PhysRevD.46.1274,Bhattacharyya2010} and a stable quark core is formed inside the neutron star.
It has also been argued that with the increase of density,
the boundaries of hadrons will gradually disappear and the transition from hadron matter to quark matter will be a smooth crossover~\cite{Baym2018RPP}.
Under this assumption, the three-window construction~\cite{2013-Masuda-ApJ,Masuda2013PTEP} was proposed.
In this paper, both of these schemes will be used.

Apart from those theoretical approaches,
on the experimental side, the study of hadron-quark phase transition relies closely on the astronomical observations of neutron stars.
Several massive neutron stars have already been detected with high-precision mass measurements~\cite{Demorest-2010_Nature467.1081,Fonseca-2016_ApJ832.167,2013-Antoniadis-Science,2020-Cromartie-NatureAstronomy,Fonseca_2021-ApJL915.L12}.
These observations provide the lower limit for the stiffness of the neutron star EOS.
The joint mass-radius observation of the neutron star from the Neutron Star Interior Composition Explorer (NICER) mission provides additional requirements~\cite{Riley_2019-ApJ887.L21,Miller2019,Riley_2021-ApJ918.L27,Miller_2021-ApJ918.L28}.
The detection of gravitational wave (GW) signals from a binary neutron star merger~\cite{Abbott2017_PRL119-161101,Abbott_2018-PhysRevLett.121.161101} has also provided an important constraint for the neutron star properties.

In previous works, numerious methods for nuclear matter and quark matter have been implemented to study the hadron-quark phase transition in the hybrid star~\cite{Schertler:1999PRC,2010-Agrawal-PhysRevD.81.023009,2013-Logoteta-PhysRevC.88.055802,2013-Orsaria-PhysRevD.87.023001,2017-Wu-PhysRevC.96.025802,2020-Miao-APJ,2021-Ju-PhysRevC.103.025809,2022-Contrera-PhysRevC.105.045808,2022-Huang-APJ,2002-Burgio-PhysRevC.66.025802,2007-KLAHN-PLB,2010-Agrawal-PhysRevD.81.023009,2015-Li-PhysRevC.91.035803,2019-Baym-APJ,2006-Nicotra-PhysRevD.74.123001,2018-Bai-PhysRevD.97.023018,2021-Bai-EPJC}.
In Refs.~\cite{2011-ChenH-PhysRevD.84.105023,2012-ChenH-PhysRevD.86.045006,2015-ChenH-PhysRevD.91.105002,2016-Yasutake-JPCS}, the BHF theory and the DSE approach have been combined. Nevertheless, this prescription is inconsistent with respect to the relativity, as the BHF theory is nonrelativistic while the DSE approach is relativistic. 
This paper utilizes the RBHF theory for nuclear matter and the DSE approach for quark matter, which are both in the relativistic framework. 
To construct the EOS of the hybrid star, the Maxwell construction and three-window construction are employed to describe the first-order phase transition and crossover, respectively. 
The mass-radius relation and tidal deformability of the hybrid star are calculated and are compared with the astronomical observations. 

This paper is organized as follows. 
In Sec.~\ref{Sec:framework}, the RBHF theory for nuclear matter, the DSE approach for quark matter, and the construction schemes for hadron-quark phase transition are briefly introduced.
The results and discussions are presented in Sec.~\ref{Sec:Result}. 
Finally, a summary is given in Sec.~\ref{Sec:Summary}. 

\section{Nuclear matter, quark matter, and hadron-quark phase transition}\label{Sec:framework}

\subsection{Nuclear matter}
In the RBHF theory, the nuclear matter is described with the nucleons as the basic degrees of freedom. 
The single-particle motion of a nucleon in nuclear medium is described by the Dirac equation
\begin{equation}\label{eqn:Dirac}
	(\bm{\alpha}\cdot\bm{p}+\beta M_\tau +\beta \hat{U}_{\tau})u_{\tau}(\bm{p},s)
	= E_{\bm{p},\tau}u_{\tau}(\bm{p},s)\,,
\end{equation}
where the subscript $\tau=n, p$ indicates neutron or proton.
$M_\tau$ is the mass of a free nucleon. 
$u_{\tau}(\bm{p},s)$ is the Dirac spinor of a nucleon with momentum $\bm{p}$, spin $s$, and single-particle energy $E_{\bm{p},\tau}$. 
The single-particle potential operator $\hat{U}_{\tau}$ is generally divided into scalar and vector components
	\begin{equation}
		\hat{U}_{\tau}(\bm{p}) =U_{S,\tau}(p) + \gamma^0 U_{0,\tau}(p)+\bm{\gamma}\cdot\hat{\bm{p}}U_{V,\tau}(p)\,.
	\end{equation}
	Here $\hat{\bm{p}}=\bm{p}/p$ is the unit vector parallel to the momentum $\bm{p}$. 
	The quantities $U_{S,\tau}(p)$, $U_{0,\tau}(p)$, and $U_{V,\tau}(p)$ are the scalar potential, the timelike part, and the spacelike part of the vector potential.
	The momentum dependent potentials can be determined uniquely by considering the positive- and negative-energy states simultaneously, i.e., the RBHF theory in the full Dirac space \cite{2021-SBWang-PRC.103.054319}.
	However, this method is now limited up to the density $\rho=0.57\ \text{fm}^{-3}$ \cite{2022-SBWang-Phys.Rev.C106.L021305}, which is not high enough to study the hadron-quark phase transition in neutron stars.
	This density limitation is absent if the momentum-independence approximation \cite{Brockmann1990} is utilized, where $U_{S,\tau}$ and $U_{0,\tau}$ are approximated to be constants and $U_{V,\tau}$ is neglected. Within this approximation, the single-particle potential operator $\hat{U}_{\tau}$ is written as
	\begin{equation}\label{eq:USU0}
		\hat{U}_{\tau} =U_{S,\tau}+\gamma^0 U_{0,\tau}\,.
	\end{equation}

By defining the effective nucleon mass $M^*_{\tau}$ and effective energy $E^*_{\bm{p},\tau}$
\begin{equation}
	M^*_{\tau}= M_\tau + U_{S,\tau}\,,\qquad E^*_{\bm{p},\tau}=E_{\bm{p},\tau}-U_{0,\tau}\,,
\end{equation}
the Dirac equation \eqref{eqn:Dirac} in nuclear medium can be rewritten as
\begin{equation}\label{eqn:Dirac-eff}
	(\bm{\alpha}\cdot\bm{p}+\beta M^*_{\tau})u_{\tau}(\bm{p},s)
	=E^*_{\bm{p},\tau}u_{\tau}(\bm{p},s).
\end{equation}
From the Dirac equation \eqref{eqn:Dirac-eff}, the dispersion relation $E^*_{\bm{p},\tau}=\sqrt{\bm{p}^2+M^{*2}_{\tau}}$ is obtained and the Dirac spinor can be solved exactly as 
\begin{equation}
	u_{\tau}(\bm{p},s) = \sqrt{\frac{E^*_{\bm{p},\tau}+M^*_{\tau}}{2M^*_{\tau}}}
	\begin{pmatrix}
		1\\
		\frac{\bm{\sigma}\cdot\bm{p}}{E^*_{\bm{p},\tau}+M^*_{\tau}}
	\end{pmatrix}
    \chi_s\chi_\tau\,,
\end{equation}
where $\chi_s$ and $\chi_\tau$ are the spin and isospin wave function respectively. 
The normalization condition is $\bar{u}_\tau(\bm{p},s)u_\tau(\bm{p},s)=1$.

In the RBHF theory, the single-particle potentials $U_{S,\tau}$ and $U_{0,\tau}$ are self-consistently determined with the effective $NN$ interaction, $G$ matrix, which can be obtained by solving the Thompson equation \cite{1970-Thompson-PhysRevD.1.110,Brockmann1990}
\begin{equation}\label{eqn:Thompson}
	G_{\tau\tau'}(\bm{q'},\bm{q}|\bm{P})
	= V_{\tau\tau'}(\bm{q'},\bm{q}) + \int\frac{d^3k}{(2\pi)^3}V_{\tau\tau'}(\bm{q'},\bm{k})
	\frac{Q_{\tau\tau'}(\bm{k},\bm{P})}{W_{\tau\tau'}-E^*_{\tau\tau'}}
	G_{\tau\tau'}(\bm{k},\bm{q}|\bm{P})\,,
\end{equation}
where $\tau\tau'=$ $nn$, $pp$, or $np$.
In Eq.~\eqref{eqn:Thompson}, $\bm{P}=(\bm{k}_1+\bm{k}_2)/2$ is the center-of-mass momentum, and $\bm{k}=(\bm{k}_1-\bm{k}_2)/2$ is the relative momentum of two interacting nucleons with momenta $\bm{k}_1$ and $\bm{k}_2$ in the rest frame of nuclear matter.
The quantities $\bm{q}$, $\bm{q'}$, and $\bm{k}$ are the initial, final, and intermediate relative momenta of the two nucleons scattering in nuclear matter, respectively.
$W_{\tau \tau'}=E^*_{\bm{P} +\bm{q},\tau}+E^*_{\bm{P}-\bm{q},\tau'}$ is the starting energy, and $E^*_{\tau\tau'}=E^*_{\bm{P}+\bm{k},\tau}+E^*_{\bm{P}-\bm{k},\tau'}$ is the total single-particle energy of intermediate two-nucleon states. 
The Pauli operator $Q_{\tau\tau'}$ avoids the $NN$ scattering to occupied states in the Fermi sea and is defined as
\begin{equation}
	Q_{\tau\tau'}(\bm{k},\bm{P})=
	\left\{
	\begin{aligned}
		&1  &\qquad &\text{if}\quad|\bm{P}+\bm{k}|>k_F^{\tau} \quad\text{and}\quad|\bm{P}-\bm{k}|>k_F^{\tau'}\,,\\
		&0  &\qquad &\text{otherwise}\,,
	\end{aligned}
	\right.
\end{equation}
where $k_F^{\tau}$ represents the Fermi momentum for nucleon $\tau$. 
For nuclear matter with total nucleon density $\rho=\rho_n+\rho_p$ and isospin asymmetry $\delta=(\rho_n-\rho_p)/(\rho_n+\rho_p)$, the Fermi momentum is calculated as $k_F^{\tau} = \left[3\pi^2(1\pm\delta)\rho/2\right]^{1/3}$.

With the $G$ matrix, one can calculate the single-particle potential energy 
\begin{equation}\label{eqn:U-2}
	U_{\tau}(p)=\sum_{s',\tau'}\int_0^{k^{\tau'}_{F}} 
	\frac{d^3p'}{(2\pi)^3} \frac{M^*_{\tau'}}{E^*_{\bm{p}',\tau'}} 
	\langle \bar{u}_\tau(\bm{p},1/2) \bar{u}_{\tau'}(\bm{p}',s')
	|\bar{G}(W)|u_\tau(\bm{p},1/2) u_{\tau'}(\bm{p}',s')\rangle\,,
\end{equation}
where $\bar{G}$ is the antisymmetried $G$ matrix and the factor $M^*_{\tau'}/E^*_{\bm{p}',\tau'}$ comes from the normalization condition above.
Alternatively, the single-particle potential energy can be obtained by sandwiching the single-particle potential operator between the Dirac spinors
\begin{equation}\label{eqn:U-1}
	U_{\tau}(p)=\frac{M^*_{\tau}}{E^*_{\bm{p},\tau}} 
	\langle \bar{u}_\tau(\bm{p},1/2) | \hat{U}_{\tau} | u_\tau(\bm{p},1/2) \rangle
	= \frac{M^*_{\tau}}{E^*_{\bm{p},\tau}}U_{S,\tau}+U_{0,\tau}\,.
\end{equation}
By combining Eqs.~\eqref{eqn:U-2} and \eqref{eqn:U-1}, one can extract the two constants $U_{S,\tau}$ and $U_{0,\tau}$ with two momenta, e.g., $0.5k^\tau_F$ and $k^\tau_F$.

Equations \eqref{eqn:Dirac}, \eqref{eqn:Thompson}, \eqref{eqn:U-2}, and \eqref{eqn:U-1} constitute a coupled set of equations that needs to be solved self-consistently.
Starting from initial values of $U^{(0)}_{S,\tau}, U^{(0)}_{0,\tau}$ in vacuum, the Dirac spinors are obtained by solving the Dirac equation \eqref{eqn:Dirac}.
Then one solves the Thompson equation \eqref{eqn:Thompson} to get the $G$ matrix and obtain the single-particle potential energy by using the integrals in Eq.~\eqref{eqn:U-2}. 
From Eq.~\eqref{eqn:U-1} a new set of values for $U^{(1)}_{S,\tau}, U^{(1)}_{0,\tau}$ are found, which are to be used in the next iteration. 
This iterative procedure is repeated until the satisfactory convergence is reached.

Once the solution is converged, the binding energy per nucleon in nuclear matter can be calculated as 
\begin{equation}
	\begin{split}
	\frac{E}{A} =& \frac{1}{\rho}\sum_{s,\tau}\int_0^{k^{\tau}_F}\frac{d^3p}{(2\pi)^3}
	\frac{M^*_{\tau}}{E^*_{\bm{p},\tau}} 
	\langle \bar{u}_{\tau}(\bm{p},s) | \gamma\cdot\bm{p}+M_\tau | u_{\tau}(\bm{p},s) \rangle \\
	&\ - \frac{1-\delta}{2}M_p - \frac{1+\delta}{2}M_n
		+ \frac{1}{2\rho}\sum_{s,s',\tau,\tau'}\int_0^{k^{\tau}_F}\frac{d^3p}{(2\pi)^3}\int_0^{k^{\tau'}_F}\frac{d^3p'}{(2\pi)^3}
	\frac{M^*_{\tau}}{E^*_{\bm{p},\tau}}\frac{M^*_{\tau'}}{E^*_{\bm{p}',\tau'}}\\
	&\ \times\langle \bar{u}_{\tau}(\bm{p},s)\bar{u}_{\tau'}(\bm{p'},s') 
	| \bar{G}(W) | u_{\tau}(\bm{p},s)u_{\tau'}(\bm{p'},s') \rangle\,. 
	\end{split}
\end{equation}

The neutron star matter is regarded as the beta-equilibrium nuclear matter consisting of protons, neutrons, electrons, and muons. 
The energy density $\varepsilon_H$ of the neutron star matter or pure hadron star is calculated as $\varepsilon_H=\rho e_{\text{tot}}$, where the total energy $e_{\text{tot}}$ is defined as
\begin{equation}
	e_{\text{tot}}(\rho,Y_n,Y_p,Y_e,Y_\mu) 
	= E/A(\rho,Y_p) + Y_pM_p + Y_n M_n + E_e/A + E_\mu /A.
\end{equation}
The quantities $E_e/A$ and $E_\mu/A$ are the contributions from electrons and muons, which are treated as gas of relativistic 
noninteracting fermions.                     
The equilibrium particle concentrations $Y_i=\rho_i/\rho\ (i=n,p,e,\mu)$ can be calculated via the $\beta$-stability condition and charge neutrality condition
\begin{subequations}
	\begin{align}
		\mu_n - \mu_p =&\ \mu_e, \\
		\mu_n - \mu_p =&\ \mu_\mu,\\
		\rho_e + \rho_\mu = &\ \rho_p,
	\end{align}
\end{subequations}
where $\mu_i\ (i=n,p,e,\mu)$ is the chemical potential for particle $i$.
For electrons and muons, the chemical potential is obtained via 
\begin{equation}
	 \mu_i = \frac{\partial e_{\text{tot}}}{\partial Y_i}.
\end{equation}
For protons and neutrons, the chemical potential is calculated as the single-particle energy at the Fermi surface. 
Once the total energy $e_{\text{tot}}$ is calculated, the pressure of the neutron star matter can be obtained as
\begin{equation}
	P_H = \rho^2 \frac{\partial e_{\text{tot}}}{\partial \rho}.
\end{equation}
Owing to the cluster effects in nuclear matter with density $\rho<0.08\ \text{fm}^{-3}$, the RBHF theory is not applicable and the EOS introduced with the Baym-Bethe-Pethick~\cite{BBP-1971_Nucl.Phys.A175.225} and Baym-Pethick-Sutherland models~\cite{BaymPethickSutherland-1971-ApJ.170.299B} is used.
It should also be noticed that the results of the RBHF calculation 
are not so smooth that numerical derivatives can be performed. Parametrization or 
regression strategies~\cite{Krastev2006, 2013-Katayama-PhysRevC.88.035805}
are often employed to the neutron star EOS. 
In this work, Gaussian process regression~\cite{2022-Huang-APJ} is implemented to process our RBHF results.

\subsection{Quark matter}
\begin{figure}[htbp]
	 	\centering
	 	\includegraphics[width=12.0cm]{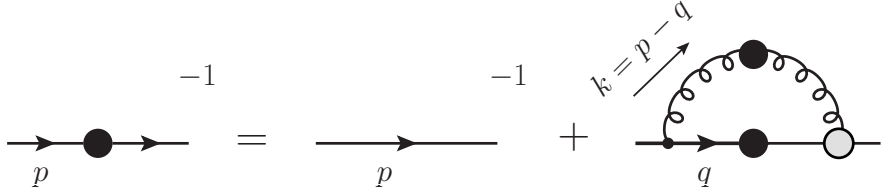}
	 	\caption{The Feynman diagram for the gap equation. 
	 	The solid (wavy) line with black thick dot is for the dressed quark (gluon) propagator, the thick gray dot represents the dressed quark-gluon interaction vertex, and the thin black dot represents the bare quark-gluon interaction vertex.}
	 	\label{fig:fey-DSE}
 \end{figure}

The DSEs are the equations of motion for fields.
They can be derived by differentiating the action of QCD and contain all the information of QCD Lagrangian. 
In this paper, the DSE of the quark propagator is concerned, the quark-gluon vertex is truncated by a symmetry preserving scheme, and the gluon propagator is approximated by the interaction model.
Therefore, the complicated DSE set can be reduced to a solvable gap equation. 
The Feynman diagram for the gap equation is shown in Fig.~\ref{fig:fey-DSE}.

At zero temperature and finite chemical potential, the gap equation for the quark propagator $S(p;\mu)$ reads as
\begin{equation}\label{eqn:DSE}
	S^{-1}(p;\mu)=Z_2(i\bm{\gamma}\cdot\bm{p}+i\gamma_4\tilde{p}_4+m_q)+\Sigma(p;\mu)\,,
\end{equation}
where $p=(\bm{p},p_4)$ is the four-momentum and $\tilde{p}_4=p_4+i\mu$ with $\mu$ is the quark chemical potential. In Eq.~\eqref{eqn:DSE}, the Euclidean metric is used, which is different from the Minkowski metric used in Eq.~\eqref{eq:USU0}.
$m_q$ is the current mass of the quark $q$ with $q=u,d,s$. In this paper, the current-quark masses $m_{u/d}=0$ are chosen for simplicity, while $m_s=115~\text{MeV}$ is obtained by fitting the $K$ meson mass in vacuum~\cite{2002-Alkofer-PhysRevD.65.094026}.
$\Sigma(p;\mu)$ is the renormalized self-energy of the quark
\begin{equation}\label{eqn:self-energy}
	\Sigma(p;\mu)= Z_1\int\frac{d^4q}{(2\pi)^4}g^2(\mu) D_{\rho\sigma}(k;\mu)  \frac{\lambda^a}{2}\gamma_{\rho}S(q,\mu) \frac{\lambda^a}{2}\Gamma_{\sigma}(q,p;\mu)\,,
\end{equation}
where $Z_1$, $Z_2$ are the renormalization constants.
$\lambda^a/2$ is the fundamental representation of $SU(3)$ color symmetry. $D_{\rho\sigma}(k;\mu)$ with $k=p-q$ is the renormalized dressed gluon propagator, and $\Gamma_{\sigma}(q,p;\mu)$ is the renormalized dressed quark-gluon vertex. 
$g(\mu)$ is the density-dependent coupling constant. 
 
To solve the gap equation~\eqref{eqn:DSE}, 
we have to know the quark-gluon vertex $\Gamma_{\sigma}(q,p;\mu)$ and gluon propagator $D_{\rho\sigma}(k;\mu)$.
In this paper, the dressed quark-gluon vertex $\Gamma_{\sigma}(q,p;\mu)$ is truncated by using the rainbow approximation~\cite{1997-Maris-PhysRevC.56.3369},
and the nonperturbative dressing effect of the dressed gluon propagator $D_{\rho\sigma}(k;\mu)$ can be absorbed in the effective interaction function $\mathcal{G}(k^2;\mu)$~\cite{2002-Alkofer-PhysRevD.65.094026}. Therefore, the interaction kernel can be expressed as
\begin{equation}\label{eqn:interaction-model}
    Z_1 g^2(\mu) D_{\rho\sigma}(k;\mu) \Gamma_{\sigma}(q,p;\mu)= \mathcal{G}(k^2;\mu)D_{\rho\sigma}^{\text{free}}(k)\gamma_{\sigma}\,,
\end{equation}
where $D_{\rho\sigma}^{\text{free}}(k)=\frac{1}{k^2}\left(\delta_{\rho\sigma}-k_{\rho}k_{\sigma}/k^2\right)$ is the free gluon propagator in the Landau gauge. 
The interaction function $\mathcal{G}(k^2;\mu)$ is usually divided into two parts: the infrared part and ultraviolet perturbative part.
At zero temperature, the quark properties are mainly determined by its infrared behavior,
so we omit the ultraviolet part as in Ref.~\cite{2011-ChenH-PhysRevD.84.105023} for better numerical behavior. 
The gluon interaction function we applied is~\cite{2011-ChenH-PhysRevD.84.105023}
\begin{equation}\label{eqn:Gauss}
	\frac{\mathcal{G}(k^2;\mu)}{k^2}=\frac{4\pi^2 D}{\omega^6} k^2 
 e^{-k^2/\omega^2}e^{-\mu^2/\omega_{\text{eff}}^2}\,.
\end{equation}
Hence, the integration is finite at the ultraviolet limit, and the renormalization procedure can be omitted by simply setting all the renormalization factors to unit.
The parameters $D$ and $\omega$ control the strength and width of the interaction in vacuum respectively. 
It is found that the observables of vector and pseudoscalar mesons are insensitive to variations of $\omega\in[0.4,0.6]$ as long as $D\omega=\text{constant}$~\cite{1999-Maris-PhysRevC.60.055214,2011-QinS-PhysRevC.84.042202}.
As in Refs.~\cite{2011-ChenH-PhysRevD.84.105023,2018-Bai-PhysRevD.97.023018}, we choose $\omega=0.5$~GeV and $D=1.0~\text{GeV}^2$. 

In the interaction model Eq.~\eqref{eqn:Gauss}, the factor $e^{-\mu^2/\omega_{\text{eff}}^2}$ depicts the effects of the medium screening on the interaction at finite chemical potential $\mu$, where $\omega_{\text{eff}}$ is the effective width of medium screening effects. 
The larger the $\omega_{\text{eff}}$, the weaker the medium screening effects and the stronger the gluon interaction. 
In Refs.~\cite{2011-ChenH-PhysRevD.84.105023,2016-ChenH-EPJA,2018-Bai-PhysRevD.97.023018,2021-Bai-EPJC}, the effective width of medium screening effects is primarily adopted as $\omega_{\text{eff}}\lesssim 0.5$~GeV to study the hadron-quark phase transition. 
In this paper, we will also study the dependence of our results on the effective width $\omega_{\text{eff}}$.

\begin{figure}[htbp]
	\centering
	\includegraphics[width=12.0cm]{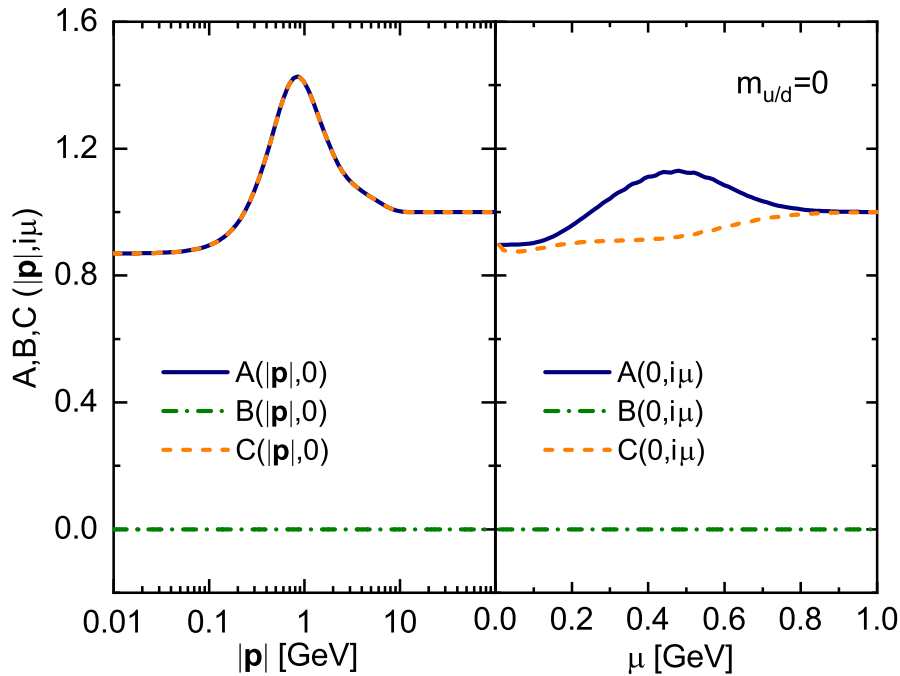}
	\caption{The Wigner solution of the quark propagator for massless quark with varying momentum (left) or varying chemical potential (right).}
	\label{fig:ABC}
\end{figure}

With the interaction model in Eqs.~\eqref{eqn:interaction-model} and \eqref{eqn:Gauss}, the self-energy in Eq.~\eqref{eqn:self-energy} can be reduced as
\begin{equation}
	\Sigma(p;\mu)= \int\frac{d^4q}{(2\pi)^4}\frac{4\pi^2 D}{\omega^6} k^2 
	e^{-k^2/\omega^2}e^{-\mu^2/\omega_{\text{eff}}^2}  \left(\delta_{\rho\sigma}-k_{\rho}k_{\sigma}/k^2\right)
	\frac{\lambda^a}{2}\gamma_{\rho}S(q,\mu) \frac{\lambda^a}{2}\gamma_{\sigma}\,.
\end{equation}
Therefore, the gap equation~\eqref{eqn:DSE} is simplified as an solvable equation in terms of the quark propagator $S(q,\mu)$, which can be decomposed according to the Lorentz structure
\begin{equation}
    S(p;\mu)^{-1} = i\bm{\gamma}\cdot\bm{p} A(\left|\bm{p}\right|^2,\tilde{p}_4^2)+B(\left|\bm{p}\right|^2,\tilde{p}_4^2)+i\gamma_4\tilde{p}_4 C(\left|\bm{p}\right|^2,\tilde{p}_4^2)\,.
\end{equation}
The scalar functions $A(\left|\bm{p}\right|^2,\tilde{p}_4^2)$, $B(\left|\bm{p}\right|^2,\tilde{p}_4^2)$, and $C(\left|\bm{p}\right|^2,\tilde{p}_4^2)$ can then be solved with the gap equation. 
It is known that the gap equation has multiple solutions. 
In vacuum, massless quarks may have the solution $B(p)\equiv0$ or $B(p) \neq 0$, which are called the Wigner solution and Nambu solution, respectively. 
The Wigner solution corresponds to the dynamical chiral symmetry (DCS) phase, where the quarks are bare and have no dynamical mass. 
The Nambu solution corresponds to the dynamical chiral symmetry breaking (DCSB) phase, where the massless quarks are dressed and the mass functions
$M(p) = B(p)/A(p)$ acquire nonzero values.
Although there are arguments that there might exist quarkyonic matter that is DCS but still confined,
it is usually believed that DCSB and confinement appear simultaneously.
Therefore, the Wigner solution corresponds to the deconfined quark phase, and the Nambu solution corresponds to the confined hadron phase.
To describe the quark core in the interior of a neutron star at high density, we only consider the Wigner solution in this paper.
In the left and right panel in Fig.~\ref{fig:ABC}, the Wigner solution of the quark propagator for the massless quark is shown as the functions of momentum and chemical potential, respectively. At zero chemical potential, the functions $A$ and $C$ are identical as the Lorentz $O(4)$ symmetry is satisfied. At zero momentum, the function $A$ shows different chemical potential dependence in comparison to the function $C$ at intermediate chemical potential, while both functions $A$ and $C$ approach unit as the quark propagator approaches asymptotic freedom at high chemical potential.

With the Wigner solution, the distribution function $f_1(\bm{p};\mu)$ can be obtained from the quark propagator 
\begin{equation}
	f_1(\bm{p};\mu)=\frac{1}{4\pi}\int_{-\infty}^{\infty}dp_4\ \text{tr}_D [-\gamma_4 S(p;\mu)]
\end{equation}
where the trace is for the spinor indices. With the quark propagator, the integration in $f_1(\bm{p};\mu)$ can be converted to a contour integral on the complex plane of $\tilde{p_4}$,
\begin{equation}
	f_1(\bm{p};\mu)
	= \frac{1}{\pi}\int_{-\infty}^{\infty}dp_4\ \frac{i(p_4+i\mu)C\left(\left|\bm{p}\right|^2,(p_4+i\mu)^2\right)}{\mathcal{M}},
\end{equation} 
where
\begin{equation}
	\mathcal{M}
	= \bm{p}^2A^2(\left|\bm{p}\right|^2,\tilde{p}_4^2) + \tilde{p}_4^2 C^2(\left|\bm{p}\right|^2,\tilde{p}_4^2) 
	+ B(\left|\bm{p}\right|^2,\tilde{p}_4^2).
\end{equation}
Then the number density $n_q$ with quark flavor $q=u,d,s$ can be obtained as a function of chemical potential~\cite{2008-ChenH-PhysRevD.78.116015}
\begin{equation}\label{eq:nqmu}
    n_q(\mu)=2N_c N_f \int\frac{d^3p}{(2\pi)^3}f_1(\bm{p};\mu)\,,
\end{equation}
where $N_f=3$ and $N_c=3$ denote the number of flavor and color, respectively.

The quark star matter is composed of quarks and leptons under the $\beta$ equilibrium and electric charge neutral condition,
\begin{subequations}\label{eq:beta&neu}
	\begin{align}
		\mu_d =&\ \mu_u+\mu_e = \mu_s\,,\\
		\mu_d =&\ \mu_u+\mu_\mu\,,\\
		\frac{2n_u-n_d-n_s}{3} =&\ n_e + n_{\mu}\,,\label{electric_charge_neutral}
	\end{align}
\end{subequations}
where the chemical potential for quark is denoted by $\mu_q$ with $q=u,d,s$.
In Eq.~\eqref{electric_charge_neutral}, the number density of lepton $n_l$ with $l=e,\mu$ is calculated as $n_l = k^3_{Fl}/3\pi^2$ where the Fermi momentum $k_{Fl}$ is related to the chemical potential $\mu_l$ by $k^2_{Fl}=\mu^2_l-m^2_l$. In the description of the pure quark star, we take $m_e=0.511~\text{MeV}$ and $m_{\mu}=105~\text{MeV}$. 
In practice, for given $\mu_n$ with
\begin{equation}\label{eq:muB}
	\mu_n = \mu_u + 2 \mu_d,
\end{equation}
one can obtain the chemical potential $\mu_i$ with $i=u,d,s,e,\mu$ by solving Eqs.~\eqref{eq:beta&neu} and \eqref{eq:muB}.

Once the quark chemical potential $\mu_q$ is obtained, the number density $n_q(\mu_q)$ can be calculated with Eq.~\eqref{eq:nqmu} and the baryon number density $\rho$ is found as $\rho = \frac{1}{3}(n_u+n_d+n_s)$. 
Furthermore, one can calculate the pressure by integrating the number density
\begin{equation}
    P_q(\mu_q)=P_q(\mu_{q,0})+\int_{\mu_{q,0}}^{\mu_q}d\mu\ n_q(\mu)\,.
\end{equation}
Theoretically, the starting point of the integral $\mu_{q,0}$ can be any value. 
In this paper we take $\mu_{u,0}=\mu_{d,0}=0$ and choose $\mu_{s,0}$ as the value of the starting point of the Wigner phase. 
Similar to Ref.~\cite{2011-ChenH-PhysRevD.84.105023}, we take the vacuum pressure $P_{u}(\mu_{u,0}) = P_{d}(\mu_{d,0}) =-45~\text{MeV}\cdot\text{fm}^{-3}$ and $P_s(\mu_{s,0}) = 0$.
Since the leptons are treated as free Fermi gas, their pressure is
\begin{equation}
		P_l = \frac{1}{24\pi^2}\left[k_{Fl} \mu_l \left(2 k_{Fl}^2 - 3 m_l^2\right) + 3 m_l^4 \ln\left(\left|\frac{k_{Fl}+\mu_l}{m_l}\right|\right) \right]\,,\quad l=e,\mu\, .
\end{equation}

Finally, with the chemical potential $\mu_i$ and number density $n_i$ with $i=u,d,s,e,\mu$, the total pressure and total energy density of the pure quark star can be obtained as
\begin{subequations}
	\begin{align}
	P_Q =& \sum_{i=u,d,s,e,\mu} P_i(\mu_i)\,,\\
	\varepsilon_Q =& \sum_{i=u,d,s,e,\mu}\mu_i n_i-P_Q\,.
	\end{align}
\end{subequations}

\subsection{Hadron-quark phase transition}

In this paper, we will study the hadron-quark phase transition as a first-order transition as well as a crossover.
The first-order phase transition is described with the Maxwell construction,
and the crossover is described with the three-window construction.
For the Maxwell construction, the phase transition occurs when the baryon chemical potential and pressure of the two phases are equal 
\begin{equation}\label{eqn:Maxwell-0}
P_H(\mu_{n,c})=P_Q(\mu_{n,c})\,,
\end{equation}
where $\mu_{n,c}$ is the critical baryon chemical potential of the first-order phase transition. 
Therefore, the pressure of the hybrid star under the Maxwell construction is
\begin{equation}\label{eqn:Maxwell-1}
P(\mu_n)=\left\{
    \begin{aligned}
        P_H\,,\qquad \text{if}\quad\mu_n\leq\mu_{n,c}\,,\\
        P_Q\,,\qquad \text{if}\quad\mu_n>\mu_{n,c}\,.
    \end{aligned}
    \right.
\end{equation}
Correspondingly, the energy density of the hybrid star under the Maxwell construction is
\begin{equation}\label{eqn:Maxwell0-2}
\varepsilon(\mu_n)=\left\{
    \begin{aligned}
        \varepsilon_H\,,\qquad \text{if}\quad\mu_n\leq\mu_{n,c}\,,\\
        \varepsilon_Q\,,\qquad \text{if}\quad\mu_n>\mu_{n,c}\,.
    \end{aligned}
    \right.
\end{equation}

For the three-window construction, a smooth interpolation of the energy density between the hadron and quark phases is performed
\begin{equation}\label{eqn:Window-0}
    \varepsilon(\rho)=f_-(\rho)\varepsilon_H(\rho)+f_+(\rho)\varepsilon_Q(\rho)\,.
\end{equation}
The interpolation function $f_\pm$ is chosen as~\cite{2013-Masuda-ApJ,Masuda2013PTEP}
\begin{equation}\label{eqn:Window-f}
	f_{\pm}=\frac{1}{2}\left(1\pm\tanh\left(\frac{\rho-\bar{\rho}}{\Gamma}\right)\right)\,,
\end{equation}
where the parameters $\bar{\rho}$ and $\Gamma$ describe the center density and the width of the transition region. 

\begin{figure}[htbp]
	\centering
	\includegraphics[width=8.0cm]{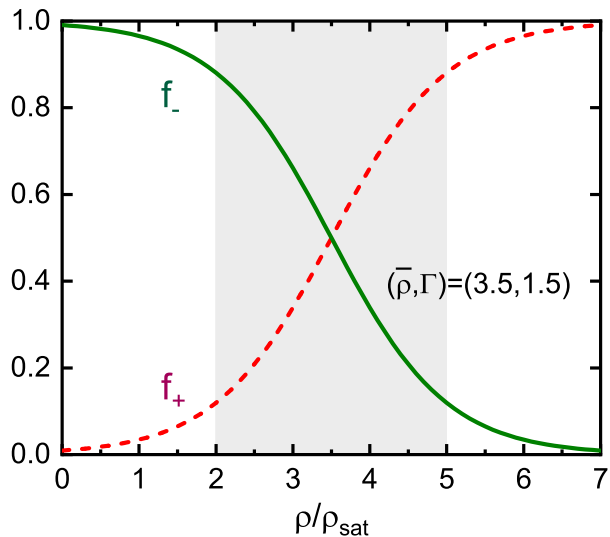}
	\caption{The interpolation functions $f_{\pm}$ with respect to the baryon density $\rho/\rho_{\text{sat}}$. The center density and the width of the transition region are $(\bar{\rho},\Gamma$)=($3.5,1.5$). The shaded region denotes the crossover region $\bar{\rho}-\Gamma\leq \rho \leq \bar{\rho}+\Gamma$.}
	\label{fig:interpolation-fun}
\end{figure}

Taking parameters ($\bar{\rho},\Gamma$)=($3.5,1.5$) as an example, the variation behavior of the function $f_{\pm}$ in terms of the baryon density is shown in Fig.~\ref{fig:interpolation-fun}.
Once the energy density of the hybrid star has been obtained, the pressure can be determined with the thermodynamic relation,
\begin{eqnarray}\label{eqn:Window-1}
	P(\rho)=\rho^2\frac{\partial(\varepsilon/\rho)}{\partial\rho}.
\end{eqnarray}

%

With the EOS of the neutron star matter, the mass-radius relation of the neutron star can be obtained by solving the Tolman–Oppenheimer–Volkov equation \cite{Oppenheimer1939_PR55-374,Tolman1939_PR55-364}, and the tidal deformability can be calculated as in Ref.~\cite{Hinderer_2010-PhysRevD.81.123016}.

\section{Results and discussion}\label{Sec:Result}

%
\begin{figure}[htbp]
	\centering
	\includegraphics[width=12.0cm]{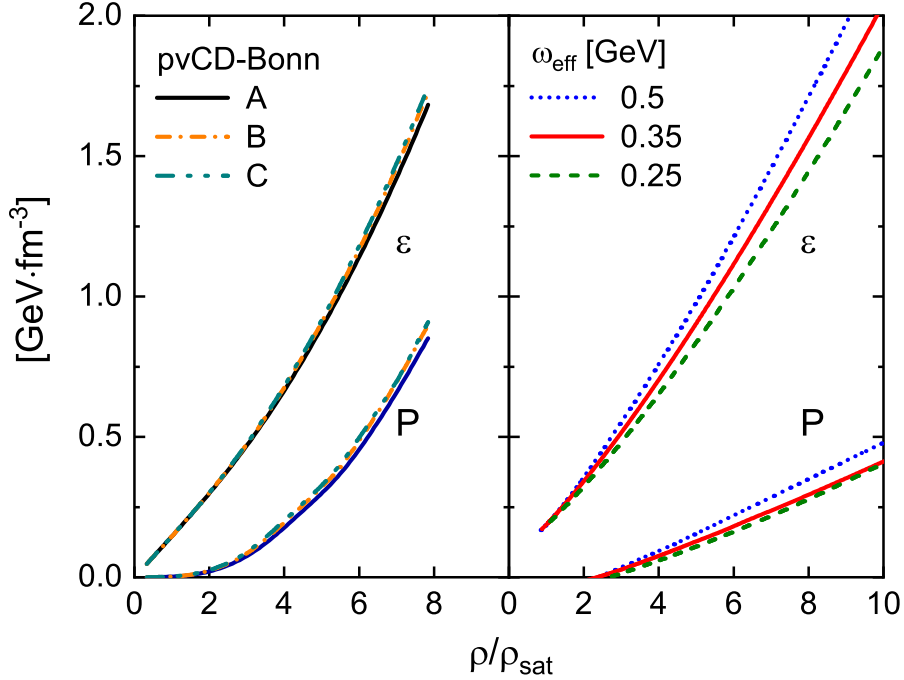}
	\caption{Left panel: energy density and pressure as functions of density for the pure hadron star
    described by the RBHF theory with potentials pvCD-Bonn A, B, and C. Right panel: similar to the left panel, but for the pure quark star described by the DSE approach with different effective width $\omega_\text{eff}$ of medium screening in the interaction model~\eqref{eqn:Gauss}.}
	\label{fig:EOS}
\end{figure}
In the left panel of Fig.~\ref{fig:EOS}, the energy density $\varepsilon$ and pressure $P$ of the pure hadron star are calculated by the RBHF theory with $NN$ interactions pvCD-Bonn A, B, and C \cite{Chencan2019CPC}.
It is clear that the difference of the results from pvCD-Bonn A, B, and C is negligible.
This is reasonable since the main difference between the three parametrizations is in the tensor force strength, which is mostly reflected in the ($T=0$) $^3S_1$-$^3D_1$ states with $T$ the total isospin.
This partial wave does not contribute to the ($T=1$) neutron-neutron state \cite{Krastev2006}, which is dominant in the neutron star.
In the following discussion, the potential pvCD-Bonn A is used, since the empirical nuclear saturation properties can be described satisfactorily by the RBHF theory with pvCD-Bonn A \cite{CCWang2020JPG}.
In the right panel, the energy density $\varepsilon$ and pressure $P$ of the pure quark star are calculated by the DSE approach with different effective width $\omega_\text{eff}$ of medium screening in the interaction model~\eqref{eqn:Gauss}.
It is found that with the effective width $\omega_\text{eff}$ increasing, i.e., the increasing of the gluon interaction, the energy density and pressure become larger.
Besides, the differences of the energy density and pressure become more evident at higher density. 

%
\begin{figure}[htbp]
	\centering
	\includegraphics[width=12.0cm]{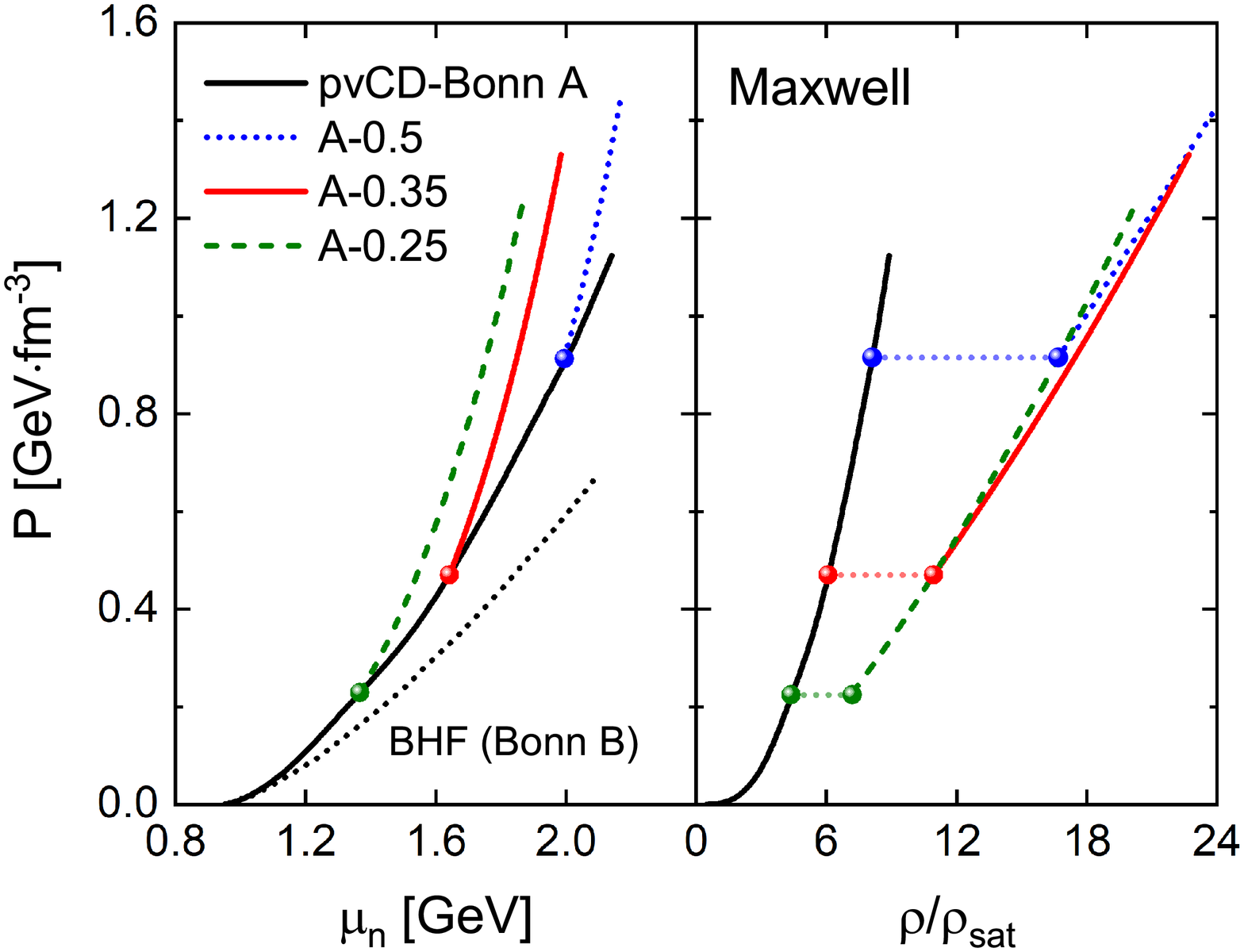}
	\caption{The $P\mbox{-}\mu_n$ relation (left) and $P\mbox{-}\rho$ relation (right) of the hybrid star under Maxwell construction with nuclear matter described by the RBHF theory with potential pvCD-Bonn A and quark matter described by the DSE approach with different effective width $\omega_\text{eff}$ of medium screening in the interaction model~\eqref{eqn:Gauss}. In comparison, the results of the hadron star described by the RBHF theory with potential pvCD-Bonn A and the $P\mbox{-}\mu$ relation obtained by the BHF theory with potential Bonn B  from Ref.~\cite{2011-ChenH-PhysRevD.84.105023} are also given.}
	\label{fig:EOS-Maxwell}
\end{figure}
%

\begin{table}[htbp]
	\centering
	\tabcolsep=0.3cm
	\caption{Critical chemical potential $\mu_{n,c}$, critical pressure $P(\mu_{n,c})$, and the phase transition density region $[\rho_H,\rho_Q]$ obtained with the RBHF theory and DSE approach under the Maxwell construction. The results obtained with the BHF theory and DSE approach from Ref.~\cite{2011-ChenH-PhysRevD.84.105023} are also shown.}
	\label{tab:tab1}
	\begin{tabular}{ccccc}
		\hline\hline
		\multirow{2}{*}{Model}  &$\mu_{n,c}$ &$P(\mu_{n,c})$ &$\rho_H$ &$\rho_Q$ \\
		&[GeV] &[GeV$\cdot \text{fm}^{-3}$] &[$\rho_{\text{sat}}$] &[$\rho_{\text{sat}}$] \\  
		\hline
            A-0.5       &1.995  &0.916  &8.105 &16.689 \\
		A-0.35      &1.640  &0.471  &6.078 &10.940 \\
		A-0.25      &1.366  &0.225  &4.379 &7.189  \\
            BHF-0.35    &1.416  &0.193  &3.556 &5.744 \\
		\hline
	\end{tabular}
\end{table}

Figure \ref{fig:EOS-Maxwell} shows the $P\mbox{-}\mu_n$ relation and $P\mbox{-}\rho$ relation of the hybrid star under Maxwell construction. 
The legend A-$\omega_\text{eff}$ denotes that the nuclear matter is described by the RBHF theory with potential pvCD-Bonn A and the quark matter is described by the DSE approach with effective width $\omega_\text{eff}$ of medium screening.
For comparison, the results of the pure hadron star described by the RBHF theory are also given. 
In the left panel, the intersection points of the EOSs in $P\mbox{-}\mu_n$ plane are the critical points of the first-order phase transition in Eq.~\eqref{eqn:Maxwell-0}.
In the right panel, the pressure is constant at the critical point of the first-order phase transition, while the density jumps from $\rho_H$ to $\rho_Q$.
The values of $\mu_{n,c}$, $P(\mu_{n,c})$, $\rho_H$, and $\rho_Q$ are listed in Table~\ref{tab:tab1}. 
It is found that for a larger $\omega_\text{eff}$, the critical baryon chemical potential $\mu_{n,c}$ and density region $[\rho_H,\rho_Q]$ are higher, and the corresponding pressure $P(\mu_{n,c})$ is larger.
For $\omega_\text{eff}=0.5$~GeV, the corresponding density of the critical point is about $8.1~\rho_\text{sat}$.
Considering that the nucleon degrees of freedom becomes less available at higher density, the parameter $\omega_\text{eff}$ should be smaller than $0.5$~GeV to obtain a reasonable EOS of the hybrid star. 
For comparison, in the left panel of Fig.~\ref{fig:EOS-Maxwell}, we also show the $P\mbox{-}\mu$ relation calculated by the BHF theory with the potential Bonn B and corresponding three-body forces from Ref.~\cite{2011-ChenH-PhysRevD.84.105023}.
The critical properties obtained with the Maxwell construction are also shown in the last row in Table~\ref{tab:tab1}.
It is found that the nonrelativistic calculations lead to a softer EOS and the first-order phase transition happens much earlier than the relativistic ones.

%
\begin{figure}[htbp]
  \centering
  \includegraphics[width=8.0cm]{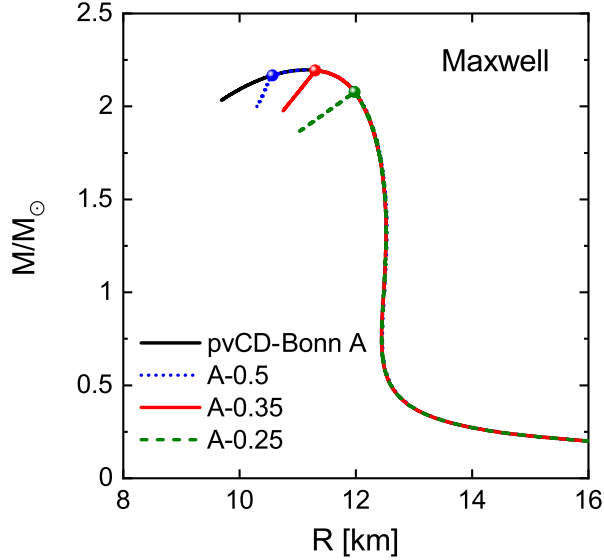}
  \caption{The mass-radius relation of the hybrid star under Maxwell construction with nuclear matter described by the RBHF theory with potential pvCD-Bonn A and quark matter described by the DSE approach with different effective width $\omega_\text{eff}$ of medium screening in the interaction model~\eqref{eqn:Gauss}. In comparison, the results of the pure hadron star described by the RBHF theory are also given.}
  \label{fig:MR-Maxwell}
\end{figure}

The mass-radius relation of the hybrid star under the Maxwell construction are plotted in Fig.~\ref{fig:MR-Maxwell}. 
It can be found that, once the first-order transition happens, the mass and radius decrease simultaneously. 
It is known that, if the mass decreases with respect to the increase of central density,
the neutron star is unstable against oscillation.
Therefore, this sharpness reflects that the appearance of quark matter leads to an unstable hybrid star.
In other words, under the Maxwell construction with the present models, there is no such a quark core in the interior of a neutron star.
We note that similar results are also found by combining the BHF theory with the DSE approach in Ref.~\cite{2011-ChenH-PhysRevD.84.105023}.

%
\begin{figure}[htbp]
	\centering
	\includegraphics[width=8.0cm]{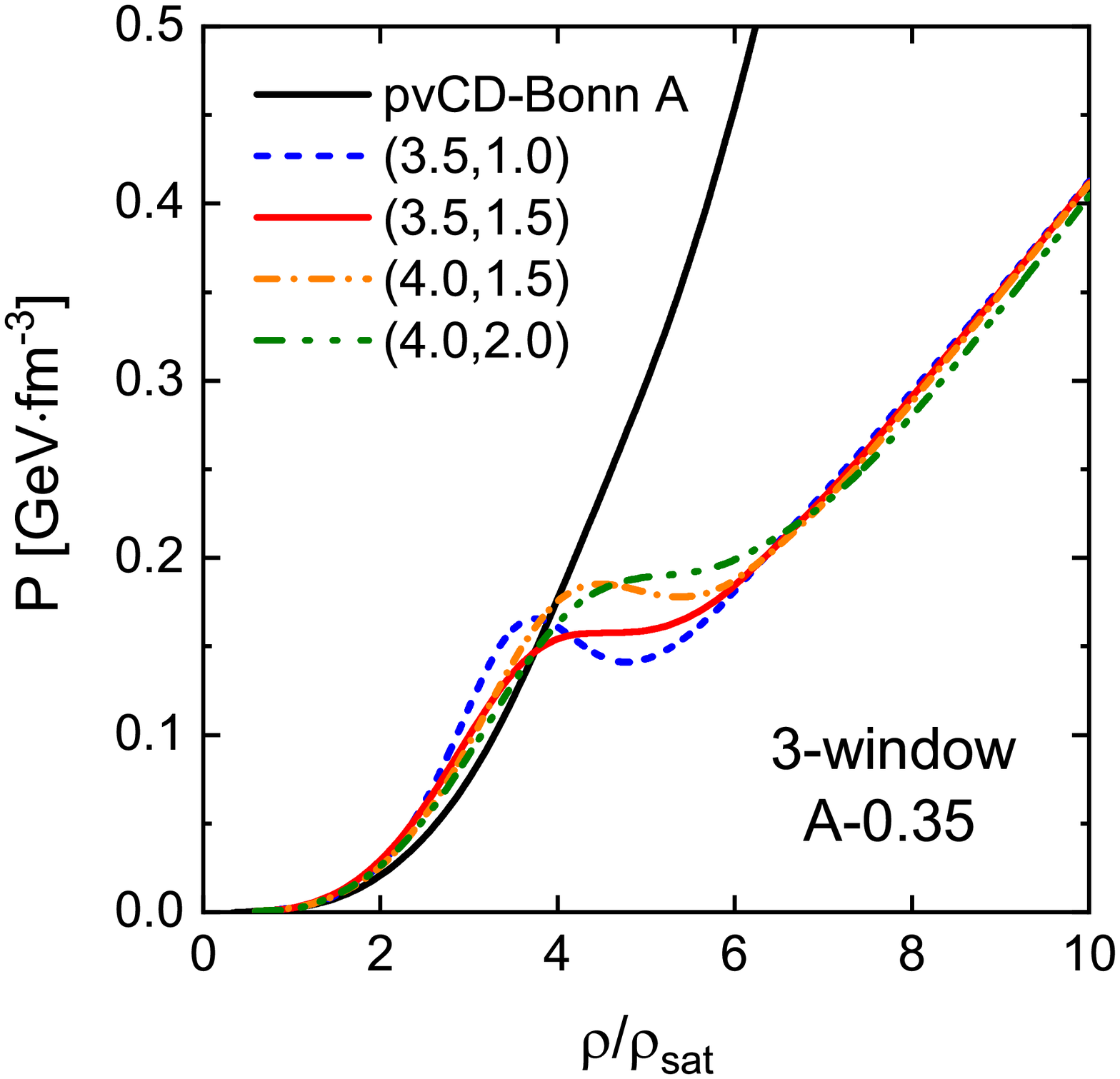}
	\caption{The pressure of the hybrid star as a function of density under the three-window construction with four representative parameter sets $(\bar{\rho},\Gamma)$.
	The results for the pure hadron star are also shown. 
    The effective width $\omega_\text{eff}=0.35$~GeV is fixed for the description of quark matter with the DSE approach.}
	\label{fig:paras-window}
\end{figure}

Under the three-window construction, a phenomenological interpolation of the energy density between the hadron and quark phases is performed, where the interpolation function is given in Eq.~\eqref{eqn:Window-f}.
The parameters $\bar{\rho}$ and $\Gamma $ characterize the center and effective width of the crossover region in baryon density.
Inspired by and basing them on Ref.~\cite{2013-Masuda-ApJ}, we consider two constraints on the choice of the two parameters:
(1) the system is always thermodynamically stable, i.e., $dP/d\rho>0$, and
(2) the crossover density region should be moderate to avoid the failure of the nuclear matter method and quark matter method, i.e., $\bar{\rho}-\Gamma>\rho_{\text{sat}}$ and $\bar{\rho}+\Gamma<6\rho_{\text{sat}}$.
In addition, a stiffer EOS is favored to satisfy the observation of massive neutron stars.

Figure \ref{fig:paras-window} depicts the $P\mbox{-}\rho$ relation of the hybrid star under the three-window construction. 
The quark matter effective width $\omega_\text{eff}=0.35$~GeV is chosen as an example and four representative parameter sets ($\bar{\rho},\Gamma $) are shown.
The parameter sets $(3.5,1.0)$ and $(4.0,1.5)$ lead to nonmonotonic pressure and should be discarded.
The maximum density of the parameter set $(4.0,2.0)$ is $6\rho_{\text{sat}}$, which is somehow too large for the RBHF theory.
The parameter set ($3.5,1.5$) satisfies the conditions (1) and (2) and is chosen in the following calculations.

%
\begin{figure}[htbp]
	\centering
	\includegraphics[width=12.0cm]{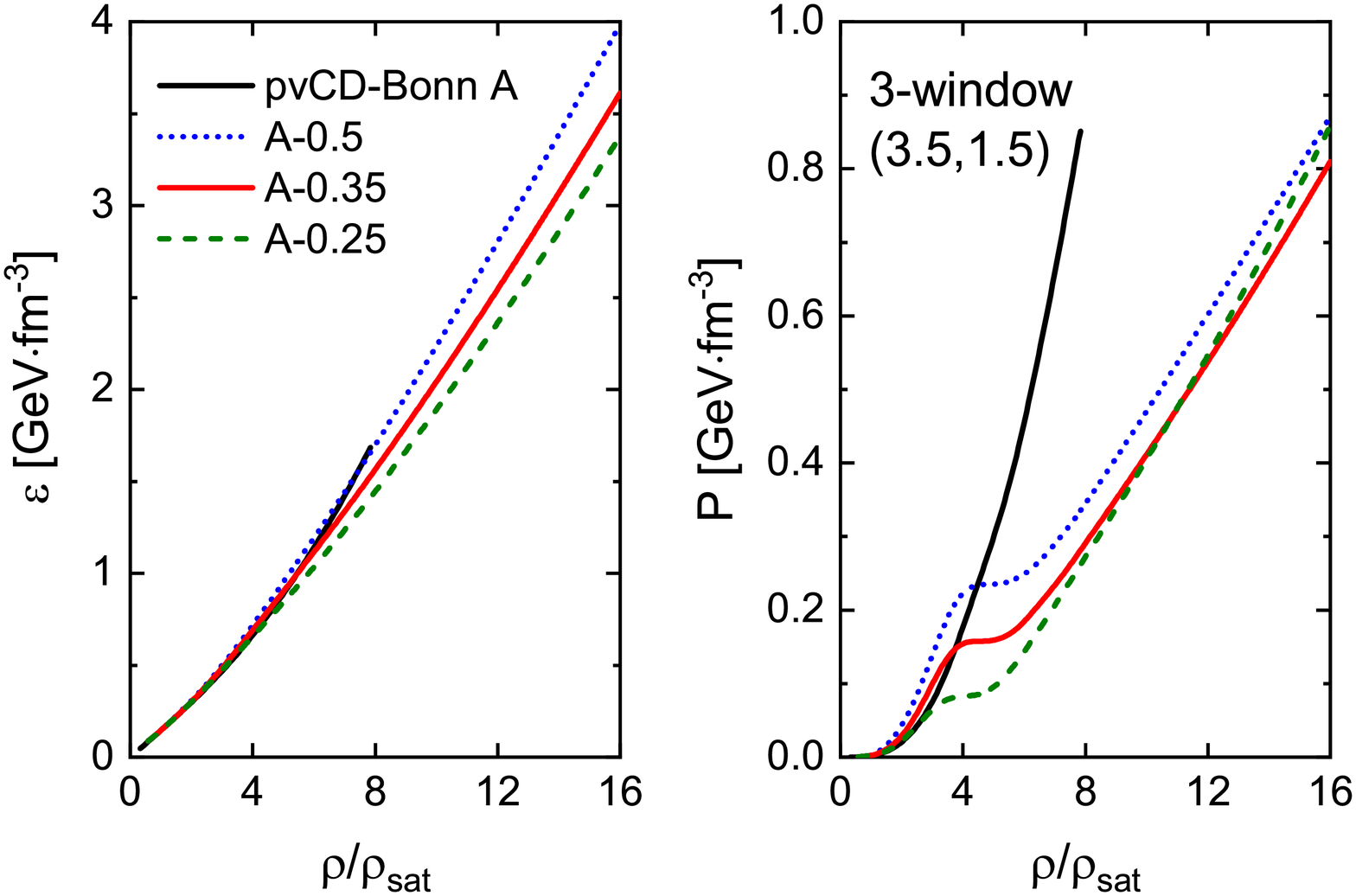}
	\caption{The $\varepsilon\mbox{-}\rho$ relation (left) and $P\mbox{-}\rho$ relation (right) of the hybrid star under the three-window construction with nuclear matter described by the RBHF theory with potential pvCD-Bonn A and quark matter described by the DSE approach with different effective width $\omega_\text{eff}$ of medium screening in the interaction model~\eqref{eqn:Gauss}. In comparison, the results of the hadron star described by the RBHF theory are also given.}
	\label{fig:EOS-Window}
\end{figure}

In Fig.~\ref{fig:EOS-Window}, the $\varepsilon\mbox{-}\rho$ relation and $P\mbox{-}\rho$ relation of the hybrid star under the three-window construction are shown. 
In the left panel, the $\varepsilon\mbox{-}\rho$ relations are interpolated in the crossover region. Outside the crossover region, pure nuclear matter or pure quark matter dominates.
In the right panel, the pressure in the crossover region increases monotonically with an increasing baryon density for the effective width $\omega_\text{eff}$ of different quarks.
From Fig.~\ref{fig:EOS-Window}, it is clear that with a lager $\omega_\text{eff}$, the energy density $\varepsilon$ and pressure $P$ are larger in the crossover region.

%
\begin{figure}[htbp]
	\centering
	\includegraphics[width=8.0cm]{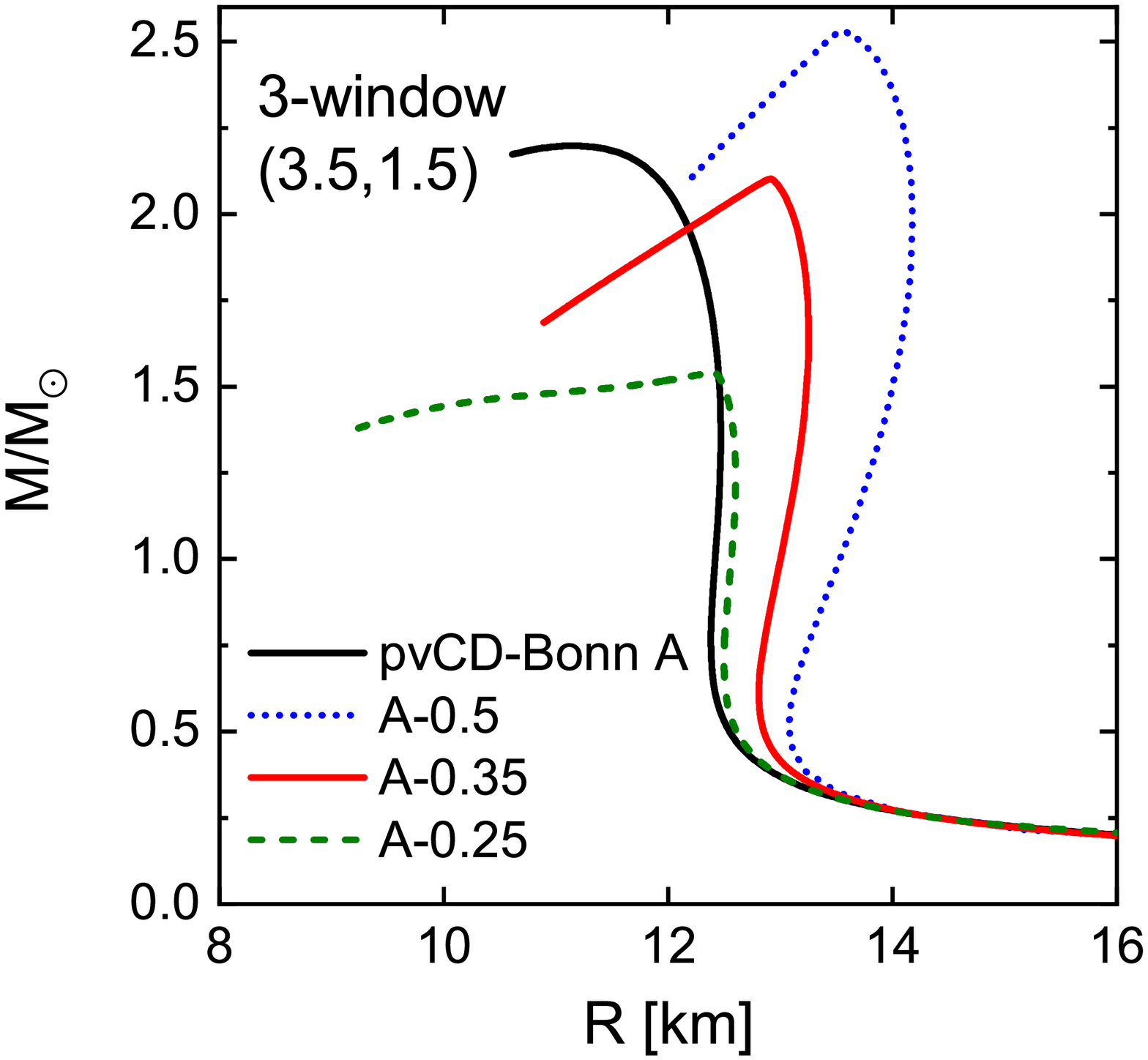}
	\caption{The mass-radius relation of the hybrid star under the three-window construction with nuclear matter described by the RBHF theory with potential pvCD-Bonn A and quark matter described by the DSE approach with different effective width $\omega_\text{eff}$ of medium screening in the interaction model~\eqref{eqn:Gauss}. In comparison, the results of the hadron star described by the RBHF theory are also given.}
	\label{fig:MR-Window}
\end{figure}

\begin{table}[htbp]
	\centering
	\tabcolsep=0.4cm
	\caption{The maximum mass $M_{\text{max}}$ of the neutron star, the corresponding radius $R_{M_{\text{max}}}$, and the radius for $1.4M_\odot$ neutron star $R_{1.4M_{\odot}}$ under the three-window construction with different effective width $\omega_\text{eff}$.}
	\label{tab:tab2}
	\begin{tabular}{cccc}
		\hline\hline
		\multirow{2}{*}{Model}&$M_{\text{max}}$ &$R_{M_{\text{max}}}$ &$R_{1.4M_{\odot}}$\\
		                      &[$M_{\odot}$] &[km] &[km]\\  
		\hline
		pvCD-Bonn A   &2.198 &11.15 &12.47\\
            A-0.5         &2.530 &13.54 &13.93\\
		A-0.35        &2.102 &12.91 &13.20\\
		A-0.25        &1.539 &12.39 &12.57\\
		\hline
	\end{tabular}
\end{table}

The mass-radius relation of the hybrid star under the three-window construction is plotted in Fig.~\ref{fig:MR-Window}. 
The maximum mass with different models, the corresponding radius, and the radius for $1.4M_\odot$ neutron star are listed in Table~\ref{tab:tab2}.
For $\omega_\text{eff}=0.25$~GeV, the maximum mass of the hybrid star is about $1.5~M_{\odot}$, which cannot support the astrophysical observation of massive neutron stars.
For $\omega_\text{eff}=0.5$~GeV, the maximum mass of the hybrid star is $2.53~M_{\odot}$.
This is in contradiction to the results from binary neutron star mergers, which require that the EOS cannot be too stiff, and provide an upper bound for the maximum mass~\cite{Annala:2017llu}. 
For $\omega_\text{eff}=0.35$~GeV, the maximum mass of the hybrid star is $2.1~M_{\odot}$, which is consistent with the current constraints from astrophysical observation. 
Therefore, to obtain a hybrid star supporting $2~M_{\odot}$ with three-window construction, the effective width $\omega_\text{eff}$ of medium screening effects should be close to $0.35$~GeV.

%
\begin{figure}[htbp]
	\centering
	\includegraphics[width=8.0cm]{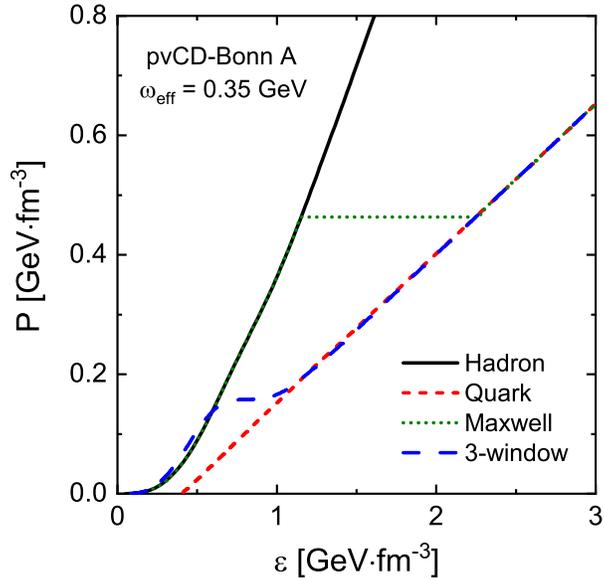}
	\caption{The $P\mbox{-}\varepsilon$ relation of the pure hadron star, the pure quark star and the hybrid star under the Maxwell construction and three-window construction. Nuclear matter is described by the RBHF theory with potential pvCD-Bonn A and quark matter is described by the DSE approach with effective width $\omega_\text{eff}=0.35$~GeV.}
	\label{fig:P-e}
\end{figure}

In Fig.~\ref{fig:P-e}, the EOSs of the pure hadron star, the pure quark star, and the hybrid star under the Maxwell construction and three-window construction are compared. 
For the Maxwell construction, there is a clear plateau of pressure as a function of energy density. This corresponds to the latent heat of the first-order phase transition. 
For the three-window construction, the central density of the phase transition region is $\bar{\rho}=3.5\rho_{\text{sat}}$ with an effective width $\Gamma=1.5\rho_{\text{sat}}$. 
Outside the crossover region, the EOS of the hybrid star asymptotically approaches that of the pure hadron star or the pure quark star. 
We also find the crossover region is lower than the first-order phase transition region in terms of energy density. 

%
\begin{figure}[htbp]
	\centering
	\includegraphics[width=8.0cm]{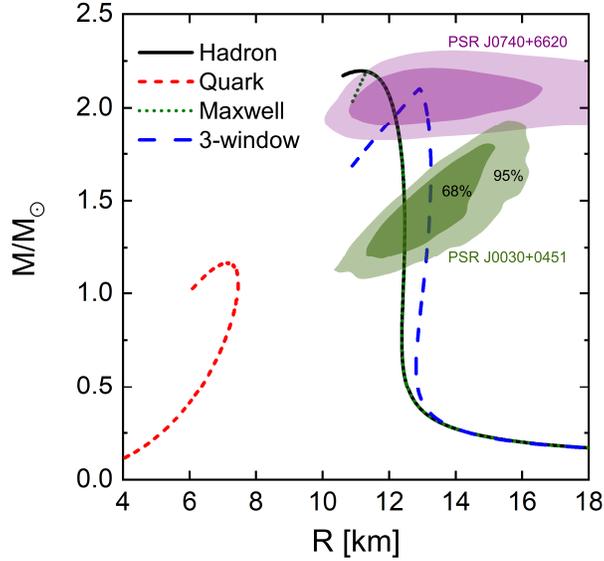}
	\caption{The mass-radius relation of the pure hadron star, the pure quark star, and the hybrid star under the Maxwell construction and three-window construction. Nuclear matter is described by the RBHF theory with potential pvCD-Bonn A and quark matter is described by the DSE approach with the effective width $\omega_\text{eff}=0.35$~GeV. The dark (light) green and purple regions indicate the $68\% (95\%)$ confidence intervals constrained by the NICER analysis of PSR J0030+0451 \cite{Miller2019} and PSR J0740+6620 \cite{Miller_2021-ApJ918.L28}.}
	\label{fig:MR}
\end{figure}

In Fig.~\ref{fig:MR}, the mass-radius relation of the pure hadron star, the pure quark star, and the hybrid star under the Maxwell construction and three-window construction is depicted.
Except for the pure quark star modeled by the DSE approach, both the pure hadron star and the hybrid star can support $2M_{\odot}$.
In comparison, the joint constraints of the mass and radius of neutron stars are also shown.
The $68\%$ and $95\%$ contours of the joint probability density distribution of the mass and radius of PSR J0030+0451 \cite{Miller2019} and PSR J0740+6620 \cite{Miller_2021-ApJ918.L28} from the NICER analysis are also shown.
It can be found that the mass radius of the pure hadron star, the hybrid stars with the Maxwell, and three-window constructions are all consistent with the recent constraints by NICER. 
The radii of a $1.4M_{\odot}$ hybrid star $R_{1.4M_{\odot}}$ under the Maxwell construction and three-window construction are $12.47$\ km and $13.20$\ km, respectively.

%
\begin{figure}[htbp]
	\centering
	\includegraphics[width=8.0cm]{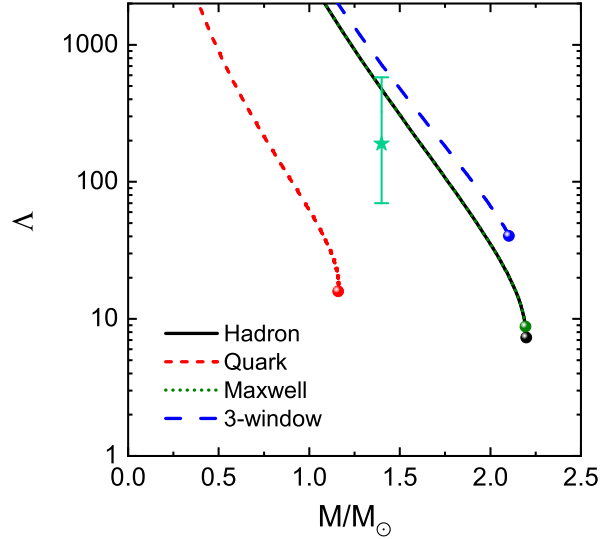}
	\caption{Tidal deformability of the pure hadron star, the pure quark star, and the hybrid star under the Maxwell construction and three-window construction. Nuclear matter is described by the RBHF theory with potential pvCD-Bonn A and quark matter is described by the DSE approach with the effective width $\omega_\text{eff}=0.35$~GeV. The constraints from GW170817 $\Lambda_{1.4M_{\odot}}=190^{+390}_{-120}$ \cite{Abbott_2018-PhysRevLett.121.161101} are also shown.}
	\label{fig:tidal}
\end{figure}

The tidal deformability of the pure hadron star, the pure quark star, and the hybrid star under the Maxwell construction and three-window construction are shown in Fig.~\ref{fig:tidal}.
The tidal deformability of the pure hadron star and that of the hybrid star under the Maxwell construction share the same value, $473$, which is consistent with the constraints $\Lambda_{1.4M_{\odot}}=190^{+390}_{-120}$ from GW170817 \cite{Abbott_2018-PhysRevLett.121.161101}.
The tidal deformability of the hybrid star under the three-window construction is $715$, which is slightly higher than that from the astronomical observation.
We notice that the RBHF theory used in the present work can be improved by considering the negative-energy states in the full Dirac space \cite{2021-SBWang-PRC.103.054319,Tong2022ApJ}, where the tidal deformability can be reduced to a value which is much closer to the center value from GW170817.
Therefore, the three-window construction with the nuclear matter described by the RBHF theory in the full Dirac space might lead to a tidal deformability which is consistent with the GW constraints. 

\section{Summary and prospects}\label{Sec:Summary}

The possible hadron-quark phase transition is explored by combining the RBHF theory for nuclear matter and the DSE approach for quark matter. 
The Maxwell construction and three-window construction are implemented to study the first-order phase transition and crossover, respectively.
For the Maxwell construction, the phase transition occurs where the baryon chemical potential and pressure of the two phases are equal.
With the RBHF theory and DSE approach, there is no stable quark core in the interior of a neutron star, which confirms previous studies with nonrelativistic Brueckner-Hartree-Fock theory and the DSE approach.
For the three-window construction, a smooth interpolation of the energy density between the hadron and quark phases is performed, where the parameters in the interpolation function are chosen in such a way as to keep the thermodynamic stability and lead to a moderate crossover density region. 
To support a two-solar-mass neutron star under the three-window construction, the effective width of medium screening effects in quark matter should be around $\omega_{\text{eff}}=0.35$~GeV.
The mass-radius relation of the hybrid star is consistent with the joint mass-radius observation, while the tidal deformability of a $1.4$ solar mass is found slightly higher than the constraints from gravitational wave detection.

In the future, this work can be extended by improving the theoretical methods for hadron matter, quark matter, and construction schemes. 
The RBHF theory can be improved by considering the negative energy states in the full Dirac space. 
For the DSE framework, the parameters cannot be changed arbitrarily as they are determined by the hadron properties as well as the phase transition at finite temperature. 
To obtain a stiffer EOS for the quark matter, it is necessary to improve the vertex and gluon truncation in the DSE approach, for example, by using the Ball-Chiu or Chang-Liu-Roberts vertex~\cite{Ball1908prd,Qin2013PLB} and including more interaction channels~\cite{Eichmann2016prd} as well as higher order corrections~\cite{Gao2021PRD}. As for the construction schemes, considering the studies by combing BHF and DSE approaches~\cite{2011-ChenH-PhysRevD.84.105023}, the Gibbs construction is likely to support a stable quark core. 
After the corrections and improvements are realized, it is hopeful to achieve a better understanding of the hadron-quark phase transition in neutron stars.
 
\begin{acknowledgments}
This work was partly supported by the China Postdoctoral Science Foundation under Grants No. 2021M700610 and No. 2022M723230; the CAS Project for Young Scientists in Basice Research (YSBR060); the National Natural Science Foundation of China under Grants No. 12147102, No. 12005060, No. 12205353, and No. 12205030; and the Fundamental Research Funds for the Central Universities under Grants No. 2020CDJQY-Z003 and No. 2021CDJZYJH-003.

\end{acknowledgments}


\bibliography{ref}
\end{document}